\documentclass[aps,twocolumn,floatfix,superscriptaddress,prx,showpacs]{revtex4-2}
\usepackage{graphicx}% Include figure files
\usepackage{epsfig}% Include figure files
\usepackage{amssymb,amsmath,amsfonts,hyperref}
\usepackage{wasysym}
\usepackage{latexsym}
\usepackage{verbatim}
\usepackage{float}
\usepackage[normalem]{ulem}
\usepackage{color}

\newcommand{\be}{\begin{eqnarray}}
\newcommand{\ee}{\end{eqnarray}}

\begin{document}
%\preprint{APS/123-QED}
%draft
%\twocolumn[\hsize\textwidth\columnwidth\hsize\csname @twocolumnfalse\endcsname
\title{Theory of collective topologically protected Majorana fermion excitations of networks of localized Majorana modes.}
\author{Kedar Damle}
\affiliation{\small{Department of Theoretical Physics, Tata Institute of Fundamental Research, Mumbai 400005, India}}
\begin{abstract}
Predictions of localized Majorana modes, and ideas for manipulating these degrees of freedom, are the two key ingredients in proposals for physical platforms for Majorana quantum computation. Several proposals envisage a scalable network of such Majorana modes coupled bilinearly to each other by quantum-mechanical mixing amplitudes. Here, we develop a theoretical framework for characterizing collective topologically protected zero-energy Majorana fermion excitations of such networks of localized Majorana modes.  A key ingredient in our work is the Gallai-Edmonds decomposition of a general graph, which we use to obtain an alternate ``local'' proof of a ``global'' result of Lov{\'a}sz and Anderson on the dimension of the topologically protected null space of {\em real skew-symmetric} (or pure-imaginary hermitean) adjacency matrices of general graphs. Our approach to Lov{\'a}sz and Anderson's result constructs a maximally-localized basis for the said null-space from the Gallai-Edmonds decomposition of the graph. Applied to the graph of the Majorana network in question, this gives a method for characterizing basis-independent properties of these collective topologically protected Majorana fermion excitations, and relating these properties to the correlation function of monomers in the ensemble of maximum matchings (maximally-packed dimer covers) of the corresponding network graph. Our approach can also be used to identify signatures of zero-energy excitations in systems modeled by a free-fermion Hamiltonian with a hopping matrix of this type; an interesting example is provided by vacancy-induced Curie tails in generalizations (on non-bipartite lattices) of Kitaev's honeycomb model. 
\iffalse
A ``Majorana zero mode'' is the technical term for a special kind of ground state degeneracy in an interacting quantum system. A single pair of such localized modes, well-separated from each other, corresponds to a two-fold degeneracy. What's special is  that this degeneracy cannot be lifted by any local perturbation acting on one of the modes. As a result, such Majorana modes can serve as particularly robust qubits in a quantum computer. Proposals for building a quantum computer using such qubits typically work with a network of such localised Majorana modes, coupled weakly to each other by quantum mechanical mixing amplitudes that generically destroy this protected degeneracy, albeit weakly. Here, we use ideas from graph theory to develop a general procedure for identifying network topologies in which this network as a whole exhibits new collective Majorana zero modes. Our theory allows us to construct the corresponding mode wavefunctions for arbitrary values of the nonzero mixing amplitudes of the network. A pair of these collective Majorana modes gives rise to a residual two-fold degeneracy of the network as a whole, and can serve as a resource for quantum computation even when the mixing amplitudes between the constituent Majorana modes of the network are strong enough to destroy all other ground state degeneracy. This precise undesrstanding of such collective Majorana modes could one day be very useful in designing Majorana-mode based quantum computers.
\fi

\end{abstract}

%\date{\today ; DRAFT VERSION }

%\pacs{71.23.-k;73.22.Pr;71.23.An;72.15.Rn}

\maketitle

\section{Introduction}
\label{sec:Introduction}

Predictions for the existence of localized Majorana fermion modes, and ideas for addressing and manipulating these degrees of freedom, are the two key ingredients in proposals for physical platforms for Majorana quantum computation. Several proposals envisage a scalable network of such Majorana modes coupled bilinearly to each other by quantum-mechanical mixing amplitudes. A key motivation for such proposals is the expectation that such a network will not be susceptible to local perturbations that can cause decoherence and degrade the performance of a quantum computer. Very roughly speaking, the idea is that a localized Majorana mode is ``one half of a  canonical fermion'', and is therefore robust to local perturbations which do not act of the ``other half'' that is spatially separated from it\cite{Stern_Lindner_review,Nayak_etal_review}.

Such a localized and topologically protected Majorana mode has been predicted~\cite{Read_Green,Ivanov} to exist in the core of vortices in a two-dimensional $p+ip$ supercondutor. In a sample with $2n$ vortices, each vortex hosts a single localized Majorana mode, resulting in a system of $2n$ modes that mix weakly with each other if the inter-vortex separations are large. A vortex-lattice state of such superconductors is thus expected to provide a natural realization of a network of Majorana modes coupled by weak mixing amplitudes that can have interesting modulations in their values~\cite{RBiswas_PRL}. The Pfaffian state at filling fraction $\nu = 5/2$  in fractional quantum hall samples is also expected~\cite{Read_Green,Ivanov} to host such Majorana modes.

%Majorana at ends of superconducting wire with spin orbit coupling.
Superconducting wires with strong spin-orbit coupling are expected to provide another
realization of Majorana modes at the ends of the wire~\cite{Kitaev_chain,Sau_etal}, including in the case with multiple channels~\cite{Potter_Lee}. These modes are expected to be relatively robust to disorder~\cite{McGinley_Knolle_Nunnenkamp}, although strong enough disorder leads to Griffiths effects that complicate the low-energy physics~\cite{Motrunich_Damle_Huse_PRB1}.
Motivated by these expectations, proposals for creating networks of coupled wires hosting such modes have also been explored~\cite{Alicea_etal}. The effect of interactions and bond disorder (disorder in the mixing amplitudes) in such networks has also been explored in some specific cases~\cite{Affleck_Rahmani_Pikulin,Li_Franz, Laumann_Ludwig_Huse_Trebst}.

Here, we develop a theoretical framework for characterizing collective topologically protected zero-energy Majorana fermion excitations of such Majorana networks in the presence of both bond disorder in the mixing amplitudes and site disorder (deleted nodes). The random deletion of nodes models local imperfections of the physical platform, which leads to a particular localized Majorana mode being absent, while the disorder in the mixing amplitudes models variations in the relative positions of the localized Majorana modes that make up the network. In this particular context, what we mean by `topologically protected'  zero-energy Majorana excitations is simply that their existence only depends on the pattern of nonzero mixing amplitudes in the network, but not on their specific nonzero values.

A key ingredient in our work is the Gallai-Edmonds decomposition~\cite{Gallai,Edmonds,Lovasz_Plummer} of a general graph, which we use to give an alternate ``local'' proof of a ``global'' result of Lov{\'a}sz and Anderson~\cite{Lovasz1,Anderson} which equates the dimension of the topologically protected null-space of a {\em real, skew-symmetric} (or pure imaginary hermitean) matrix to the number of monomers in any maximum matching (maximally-packed dimer configuration) of the associated graph~\cite{footnote1,Rabin_Vazirani}. 
Our proof of this result
%of Lov{\'a}sz and Anderson~\cite{footnote1,Rabin_Vazirani} 
uses the Gallai-Edmonds decomposition of the associated graph to construct a maximally-localized basis for the said null-space~\cite{footnote2,Cherian,Kececioglu_Pecqueur}. A key aspect of our construction is that the topologically protected zero modes obtained in this way depend only on the pattern of the nonzero connections in the network, and are determined without any reference whatsoever to the other nonzero eigenvalues that make up the rest of the spectrum.

This construction, which is an essential part of our proof of the Lov{\'asz}-Anderson result,
%of a maximally-localized basis for the topologically protected null-space of the skey-symmetric adjacency matrix of a general graph 
can be viewed as a natural generalization of the local argument given in Ref.~\cite{Bhola_etal_PRX} for the corresponding results of Longuet-Higgins~\cite{Longuet-Higgins} on bipartite hermitean matrices (without any requirement of skew-symmetry). Indeed, Ref.~\cite{Bhola_etal_PRX} drew on an earlier proposal~\cite{Sanyal_Damle_Motrunich_PRL} for the origin of such topologically protected zero-energy eigenstates in the tight-binding model for diluted graphene and used the bipartite version of the Gallai-Edmonds decomposition (the Dulmage-Mendelsohn decomposition~\cite{Lovasz_Plummer,Dulmage_Mendelsohn,Kavitha}) to construct a maximally-localized basis for such topologically protected zero modes of bipartite hermitean matrices. Our construction here generalizes this analysis to hermitean skew-symmetric matrices {\em without any requirement of a bipartite structure}. 

Thus, Ref.~\cite{Bhola_etal_PRX} characterizes topologically protected zero modes of hopping Hamiltonians in which the hopping amplitudes can be arbitrary complex numbers, but the hopping only connects nearest-neighbour sites of a bipartite lattice. In contrast, the present analysis characterizes such topologically protected zero modes of hopping Hamiltonians in which the hopping amplitudes are purely imaginary, but the hopping can connect any two vertices of a completely general graph; this includes for instance examples such as the honeycomb and square lattices with arbitrary further neighbour hopping, so long as all hopping amplitudes are purely imaginary.

Applied to the Majorana network Hamiltonian, this gives a method for characterizing basis-independent properties of collective topologically protected zero-energy Majorana fermion excitations of the network, and relating these properties to the correlation function of monomers in the maximally-packed dimer model on the corresponding network graph. In particular, it provides convenient access to basis-independent localization properties of the zero-energy on-shell Green function of the network.  
This is clearly of significance in the context of various proposals for physical platforms for Majorana computation, since these collective topologically protected Majorana excitations of the network as a whole can serve as qubits in a Majorana quantum computer even when the original Majorana qubits of the network are destroyed by strong mixing amplitudes. Our approach can also be used to identify signatures of zero-energy excitations in other systems modeled by a free-fermion Hamiltonian with a hopping matrix of this type. In particular, it can be used to understand vacancy-induced Curie tails in Kitaev-type models of Majorana spin liquids and mean field theories for such a spin liquid state~\cite{Kitaev_anyon,Yao_Lee,Yao_Kivelson_PRL,Chua_Yao_Fiete_PRB,Lai_Motrunich_PRB,MotomehypononagonKitaev,Sanyal_etal_SU2,JFu_PRB,Biswas_Fu_Laumann_Sachdev}

%and Edmonds~\cite{Edmonds_bipartite} 

%For the present case of a general graph and its skew-symmetric adjacency matrix, no analogous result appears to have been formulated in the graph theory literature to date. 

The remainder of this artice is organized as follows: In Sec.~\ref{sec:Majorananetwork}, we introduce the Majorana network Hamiltonian of interest to us, 
%and review the usual procedure for obtaining the eigenstates of such systems. In Sec.~\ref{sec:Nullvectors}, 
summarize the classical results of Tutte, Lov{\'a}sz, and Anderson on the dimension of the null space of real skew-symmetric (equivalently, pure imaginary hermitean) matrices, and review how they determine the number of topologically protected zero-energy Majorana excitations of the network as a whole. Sec.~\ref{sec:Lovaszlocal} is devoted to our construction of a basis of maximally-localized wavefunctions for the collective topologically protected Majorana fermion excitations of the network. In Sec.~\ref{sec:Algorithmic}, we discuss some algorithmic issues that are important in the limit of very large Majorana networks.  In Sec.~\ref{sec:Discussion} we discuss the applicability of our approach to quadratic canonical fermion systems, outline its connection to earlier results, give some simple examples of our construction at work, and sketch some possibilities for further work.

\iffalse
Here, we develop a theoretical framework for characterizing {\em emergent} zero-energy Majorana fermion excitations of such networks of localized Majorana modes. A key ingredient in our work is the Gallai-Edmonds decomposition of a general graph, which we use to obtain alternate ``local'' proof of a ``global'' theorem of Lov{\'a}sz on the dimension of the null-space of skew-symmetric adjacency matrices of general graphs. Our approach yields a maximally-localized basis for this null-space, which allows us to characterize basis-independent aspects of the spatial structure of these emergent Majorana fermion excitations by relating them to the spatial distribution of monomers in the ensemble of maximum matchings of corresponding network graph. 
\fi

\iffalse
\begin{figure}
\includegraphic
\fi

\section{Counting zero-energy eigenstates of a Majorana network}
\label{sec:Majorananetwork}
As envisaged in various proposals for Majorana quantum computation, we consider a system of localized Majorana modes labeled by their position $r$, with corresponding Majorana fermion operators $\eta_r$ satisfying the anticommutation relations
\begin{eqnarray}
\{ \eta_r, \eta_{r'} \} = 2 \delta_{r r'}
\label{eq:anticommutation}
\end{eqnarray}
 These modes are coupled bilinearly by quantum-mechanical mixing amplitudes, leading to a many-body Hamiltonian of the form
\begin{eqnarray}
H_{\rm Majorana} &=& \frac{i}{4} \sum_{r r'} a_{r r'} \eta_{r} \eta_{r'} \; ,
\label{eq:H_Majorana}
\end{eqnarray}
where $a_{r r'} = - a_{r' r}$ are real-valued amplitudes that couple modes at sites $r$ and $r'$, and the factor of $4$ is merely a matter of convention.
As already noted, the quadratic Hamiltonian $H_{\rm Majorana}$ is also of interest in the context of various generalizations of Kitaev's honeycomb model to other three-coordinated lattices~\cite{Kitaev_anyon,Yao_Lee,Yao_Kivelson_PRL,Chua_Yao_Fiete_PRB,Lai_Motrunich_PRB,MotomehypononagonKitaev,Sanyal_etal_SU2,JFu_PRB,Biswas_Fu_Laumann_Sachdev}.

\subsection{Basic formalism}
\label{subsec:basicformalism}

The excitation spectrum of $H_{\rm Majorana}$ follows immediately from the eigenspectrum of the $2L \times 2L$ skew-symmetric hermitean matrix $ia_{r r'}$ in the usual way, which we recap here for completeness. The nonzero eigenvalues of $ia_{r r'}$ come in pairs that have a common magnitude, with the eigenvectors being complex conjugates of each other. Denote these pairs as ($\epsilon_{p}$, $-\epsilon_p$) where $p = 1,2\dots N$ ($N\leq L$) and $\epsilon_p > 0$ by definition, and let ($\psi_{p}(r)$, $\psi^{*}_{p}(r)$) be the corresponding eigenvectors. In addition, the eigenvalue $\epsilon=0$ can occur with an even multiplicity $2Z$ (with $Z=L-N$). We denote the corresponding null eigenvectors by $\phi_q(r)$ with $q=1,2,\dots 2Z$; note that the wavefunction amplitudes $\phi_q(r)$ can be chosen to be purely real, unlike the wavefunctions $\psi_p(r)$ which are complex valued in general.

With this in hand, we define $N$ canonical fermion operators 
\begin{eqnarray}
f_{p} &=& \frac{1}{\sqrt{2}}\sum_r \psi_{p}^{*}(r) \eta_r \; \; {\rm for} \; \;   (p = 1,2 \dots N)\; , \nonumber \\
&&
\end{eqnarray}
and the $2Z$ Majorana operators
\begin{eqnarray}
\gamma_q &=& \sum_r \phi_{q}(r) \eta_r \;\; {\rm for} \; \; (q=1,2 \dots 2Z) \; . \nonumber \\
&& \; .
\end{eqnarray}
As will be clear presently, these correspond to the $2Z$ zero-energy Majorana fermion excitations
of $H_{\rm Majorana}$.  Using the orthonormality of the eigenvectors of the skew-symmetric hermitean matrix $ia_{r r'}$, it is easy to verify that the $2Z$ Majorana operators and the $N$ canonical fermions obey the standard anticommutation relations:
\begin{eqnarray}
\{ f_{p}, f^{\dagger}_{p'} \} &=& \delta_{p p'} \nonumber \; , \\
\{ \gamma_{q}, \gamma_{q'} \} &=& 2 \delta_{q q'} \; ,
\end{eqnarray}
with all other anticommutators being identically zero. 
If we form (arbitrarily-chosen) pairs ($\gamma_{q}$, $\gamma_{\bar{q}}$), say with $q=1,2,\dots Z$ and $\bar{q} = q+Z$, we can also define $Z$ canonical fermion operators and complex wavefunctions based on this pairing
\begin{eqnarray}
g_{q} &=& \frac{1}{2} \left(\gamma_{q} +i \gamma_{\bar{q}} \right)  \; \; {\rm for} \; \; (q = 1,2 \dots Z) \;  , \nonumber \\
\zeta_q (r) & = & \phi_q(r)  - i \phi_{\bar{q}}(r) \; .\nonumber \\
&&
\end{eqnarray}
This is sometimes convenient for representing the zero mode contribution to various quantities.

Using the resolution of identity
\begin{eqnarray}
\delta_{r r'} &=& \sum_{q=1}^{2Z} \phi_q(r)\phi_q(r') + \sum_{p=1}^{N} (\psi_{p}^{*}(r) \psi_{p}(r') + {\rm h.c.}) \; ,
\end{eqnarray}
we can write express all the $\eta_r$ in terms of terms of $f_{p}$ and $g_q$ as
\begin{eqnarray}
\eta_r &=& 
\sum_{q=1}^{Z}\left( \zeta_{q}(r) g_{q} + \zeta_{q}^{*}(r) g_{q}^{\dagger}\right) \nonumber \\
&&+ \sqrt{2}\sum_{p=1}^{N}\left( \psi_{p}(r) f_{p} + \psi_{p}^{*}(r)f_{p}^{\dagger} \right)  \; . \nonumber \\
&&
\end{eqnarray}

Upon using this expression for $\eta_r$, we find
\begin{eqnarray}
H_{\rm Majorana} &=& \sum_{p=1}^{N} \epsilon_{p} \left( f^{\dagger}_{p} f_{p} - 1/2 \right) \; .
\end{eqnarray}
Thus, the canonical fermions $f_{p}$ correspond to the nonzero energy excitations of the Majorana network, while the Majorana operators $\gamma_q$ determine the ground-state degeneracy of
$H_{\rm Majorana}$ (since the occupation state of the canonical orbitals $g_q$ constructed from pairs of these Majorana operators do not enter the expression for the energy).

For the purposes of this article, the basic message from the elementary analysis reviewed here is that the wavefunctions $\phi_q(r)$ of the null vectors of the skew-symmetric hermitean matrix $ia_{r r'}$ determine the spatial structure of the collective zero-energy Majorana excitations of the network as a whole. With this background, our goal in what follows is to develop a general construction of a maximally-localized basis for the topologically protected null space of $ia_{r r'}$.

\subsection{Matchings and Tutte's Theorem}
\label{subsec:Tutte}
A key result of Tutte~\cite{Tutte1,Tutte2} on the absence topologically protected null vectors of $a_{r r'}$ is based on viewing $a_{r r'}$ as the real-valued skew-symmetric adjacency matrix of an associated graph. Since Tutte's argument involves the combinatorial problem of maximum matchings on this graph, we begin with a quick digression that reviews the relevant concepts and terminology; this will also come in handy in our subsequent discussion of the Gallai-Edmonds decomposition of a general graph in Sec.~\ref{sec:Lovaszlocal}.

\subsubsection{Matchings}

A {\em matching}  of a graph pairs up each vertex of the graph with an adjacent vertex, with the constraint that two distinct vertices do not have the same partner. If we put dimers (hard rods) down on each edge (nearest-neighbour link)  that connects a vertex with its partner, we can represent a matching as a configuration of a lattice gas of dimers that obey a hard-core constraint (no two dimers can touch at a vertex). If all vertices are matched, one has a {\em perfect matching}, or a {\em fully-packed dimer cover}. As a result of the hard-core constraint, a perfect matching may not be possible and some minimum number of vertices may have to be left unmatched; these are said to host {\em monomers} of the corresponding dimer cover. In such cases, it is interesting to consider {\em maximum matchings} ({\em maximally-packed dimer covers}), {\em i.e.} matchings that match the maximum possible number of vertices of a graph and have the largest possible number of dimers. 

Given a particular matching of a graph, an {\em alternating path} is defined as a path that starts at some vertex and goes alternately along matched and unmatched edges of the graph, with no vertex repeated and no edges traversed twice. If an alternating path returns to the starting vertex, it is an alternating cycle. Thus, in an alternating cycle, each vertex has one matched and one unmatched edge of the cycle incident upon it. With this terminology in hand, it is easy to see that a matching is a maximum matching if and only if there are no alternating paths of odd length (odd number of edges) connecting two unmatched sites.

\subsubsection{Tutte's Theorem}
We can now state Tutte's theorem using this terminology:
A graph has a perfect matching (fully-packed dimer cover) if and only if its skew-symmetric adjacency matrix has no topologically protected null vectors. An elementary proof,  relying only on the definition of a determinant and the cycle decomposition of permutations, goes as follows. Consider the determinant of $a_{rr'}$ defined as a sum over permutations
\begin{eqnarray}
{\rm det}(a) = \sum_{P} {\rm sgn}(P) \prod_{r=1}^{2N} a_{r\;P(r)} \; ,
\label{expansionofdet}
\end{eqnarray}
and decompose $P$ into its cycles. If $P$ has a one-cycle, {\em i.e.} it leaves any $r$ fixed, then
the corresponding product in Eq.~\ref{expansionofdet} has a diagonal element as one factor, and must be zero by skew-symmetry of $a_{rr'}$. Next, consider a permutation $P$ that has some cycles of odd length greater than one (involving three or more distinct vertices) in its decomposition. For each such cycle $c$ involving an odd number of elements, there is another permutation $\tilde{P}_{c}$ that differs from $P$ only by the replacement of $c$ by $c^{-1}$; all other cycles of $P$ remain unchanged in going from $P$ to $\tilde{P}_c$. By the skew-symmetry of $a_{r r'}$, the contributions of $P$ and $\tilde{P}_c$ cancel each other in the expansion of the determinant. This is because the product associated with $\tilde{P}_c$ in Eq.~\ref{expansionofdet} contains an odd number of matrix elements whose sign has been reversed relative to the corresponding factors in the product associated with $P$ (due to the skew-symmetry of $a_{rr'}$), while ${\rm sgn}(P) = {\rm sgn}(\tilde{P}_c)$. Therefore, due to such cancellations, ${\rm det}(a)$ for a skew-symmetric matrix can be written as a restricted sum over permutations that have no cycles with an odd number of distinct vertices:
\begin{eqnarray}
{\rm det}(a) = \tilde{\sum}_{P} {\rm sgn}(P) \prod_{r=1}^{2L} a_{r\;P(r)} \; ,
\label{restrictedexpansionofdet}
\end{eqnarray}
where the tilde on the sum reminds us that it is taken over permutations whose cycle decomposition only has cycles involving an even number of elements.

Next, notice that any term in the restricted expansion (Eq.~\ref{restrictedexpansionofdet}) of the determinant corresponds to $2^{l_P}$ different perfect matchings of the associated graph, where $l_P$ is the number of independent cycles of four or more elements in the decomposition of the permutation $P$. This is because each even cycle in $P$ corresponds to a simple non-intersecting loop of even length on the associated graph, and such even length loops can always be perfectly matched. In fact, any such loop of length four or more  can be perfectly matched by dimers in {\em two} different ways, while a length two loop involving just two vertices is just a `doubled' edge (traversed in both directions), which has a unique perfect matching obtained by placing a dimer on this edge. 

Therefore, if ${\rm det}(a)$ does {\em not} vanish {\em identically}, {\em i.e.}, does not vanish as an algebraic expression written in terms of matrix elements treated as independent variables, the associated graph {\em must} admit a perfect matching. Conversely, if the associated graph has a perfect matching, then the expansion of ${\rm det}(a)$ has at least one term that is not identically zero. Thus, a perfect matching exists if and only if ${\rm det}(a)$ does not vanish identically.
Now, note that $a_{r r'}$ has a topologically protected null vector if and only if ${\rm det} (a)$ vanishes identically as an algebraic expression written in terms of matrix elements treated as independent variables.  Therefore, we have proved Tutte's result: that ${\rm det}(a)$ is nonzero if and only if $a_{r r'}$ has no topologically protected null vectors.

\subsection{Lov{\'a}sz and Anderson's generalization of Tutte's theorem}
\label{subsec:Lovaszglobal}
Since Tutte's theorem relates the absence of any topologically protected null vectors of $a_{r r'}$ to the existence of a perfect matching of the associated graph, it is natural to ask if the dimension of the topologically protected null space of a real skew-symmetric matrix $a_{r r'}$ is related in a simple way to the number of monomers in any maximum matching of the associated graph. Lov{\'a}sz~\cite{Lovasz1} and Anderson~\cite{Anderson} proved that the two are indeed precisely equal to each other. Here, we follow the exposition of Rabin and Vazirani~\cite{Rabin_Vazirani} and sketch a version of the proof which relies on a classical result of Frobenius on antisymmetric matrices~\cite{Kowaleski}.

Consider the graph corresponding to a $2L \times 2L$ skew-symmetric real matrix $a_{r r'}$. Let $2Z$ be the number of topologically protected null vectors of $a_{r r'}$, and let $N \equiv L-Z$ as before.
Let any maximum matching of the associated graph have $2M$ monomers in it; {\em i.e.} it leaves $2M$ vertices of this graph uncovered. Consider any one maximum matching of this graph, and consider the subgraph defined by the matched vertices of this maximum matching and edges between them. Since this subgraph is perfectly matched, Tutte's theorem tells us that the corresponding submatrix of $a_{r r'}$ has no topologically protected null vectors. Equivalently, its determinant does not vanish identically, {\em i.e.} there is at least one choice of values of the nonzero matrix elements that makes its rank equal to $2L-2M$. 

This implies $2Z \leq 2M$. To see why, assume that this is not true, and $2Z> 2M$. Then, by the rank-nullity theorem, any submatrix of $a_{r r'}$ of size greater than $2L-2Z$ must have a determinant that vanishes identically. Since we have produced a submatrix of size $2L-2M$ which has a determinant that does not vanish identically, we must have $2L-2M \leq 2L - 2Z$, implying 
\begin{eqnarray}
Z &\leq & M \; .
\label{eq:inequality1}
\end{eqnarray}
\begin{figure}
 		\includegraphics[width=\columnwidth]{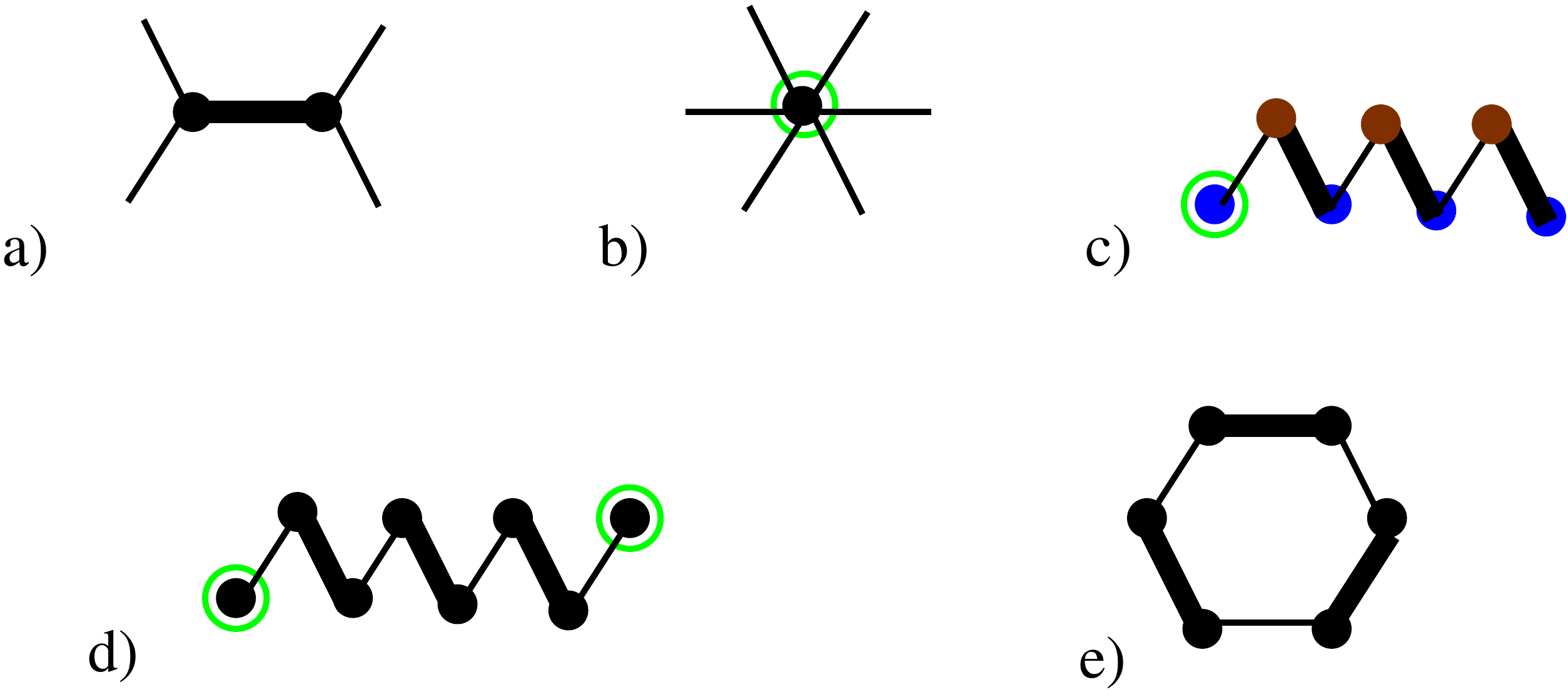}
\caption{a) A pair of adjacent vertices {\em matched} to each other, {\em i.e.} connected by a dimer on the link between them. b) An unmatched vertex hosting a monomer, {\em i.e.} not connected by a dimer to any of its neighbors. c) A segment of an {\em alternating path} starting from a monomer. d) An {\em augmenting path}, {\em i.e.} an alternating path that starts and ends at a monomer. e) An alternating cycle. See Sec.~\ref{sec:Lovaszlocal} for a detailed discussion.}
\label{fig:MatchingTerminology}
 \end{figure}
 
To establish the reverse inequality, consider a specific assignment of values to the nonzero elements of $a_{r r'}$ which ensures that the resulting matrix has exactly $2Z$ null vectors.  
By the rank-nullity theorem, there is exists a $2N \times 2N$ nonsingular submatrix $A_{\alpha \beta}$ of $a_{r r'}$ such that the row indices of elements of this submatrix form the subset $\alpha$ of $2N$ distinct indices drawn from $(1,2,...2L)$, while the column indices of elements of this submatrix form another subset $\beta$ of $2N$ distinct indices drawn from $(1,2,..2L)$. These two subsets $\alpha$ and $\beta$ are identical only if the
submatrix $A_{\alpha \beta}$ is {\em symmetrically located}, which is not guaranteed in general.

However, for our {\em skew-symmetric} matrix $a_{r r'}$, one can appeal to a theorem of Frobenius~\cite{Kowaleski}, which guarantees that ${\rm det}(A_{\alpha \alpha}) {\rm det}(A_{\beta \beta}) = (-1)^{2N}{\rm det}(A_{\alpha \beta})$ (using the notation introduced above). Since ${\rm det}(A_{\alpha \beta})$ is nonzero, this guarantees the existence of a symmetrically-located nonsingular $2N \times 2N$ submatrix $A_{\alpha \alpha}$ with nonzero determinant. This symmetrically-located submatrix $A_{\alpha \alpha}$ of $a_{r r'}$ can be viewed as the skew-symmetric adjacency matrix of the subgraph consisting of the subset $\alpha$ of vertices, with edges corresponding to nonzero elements of $A_{\alpha \alpha}$. Since $A_{\alpha \alpha}$ has nonzero determinant, this subgraph has a perfect matching by Tutte's theorem. Therefore, the size $2L-2M$ of the maximum matching of the original graph must satisfy $2L-2M \geq 2N$, {\em i.e.} $2L-2M \geq 2L-2Z$. In other words
\begin{eqnarray} 
Z &\geq& M \; .
\label{eq:inequality2}
\end{eqnarray}
Combining these two inequalities Eq.~\ref{eq:inequality1} and Eq.~\ref{eq:inequality2}, we have proved $Z=M$.
\begin{figure}
 		\includegraphics[width=\columnwidth]{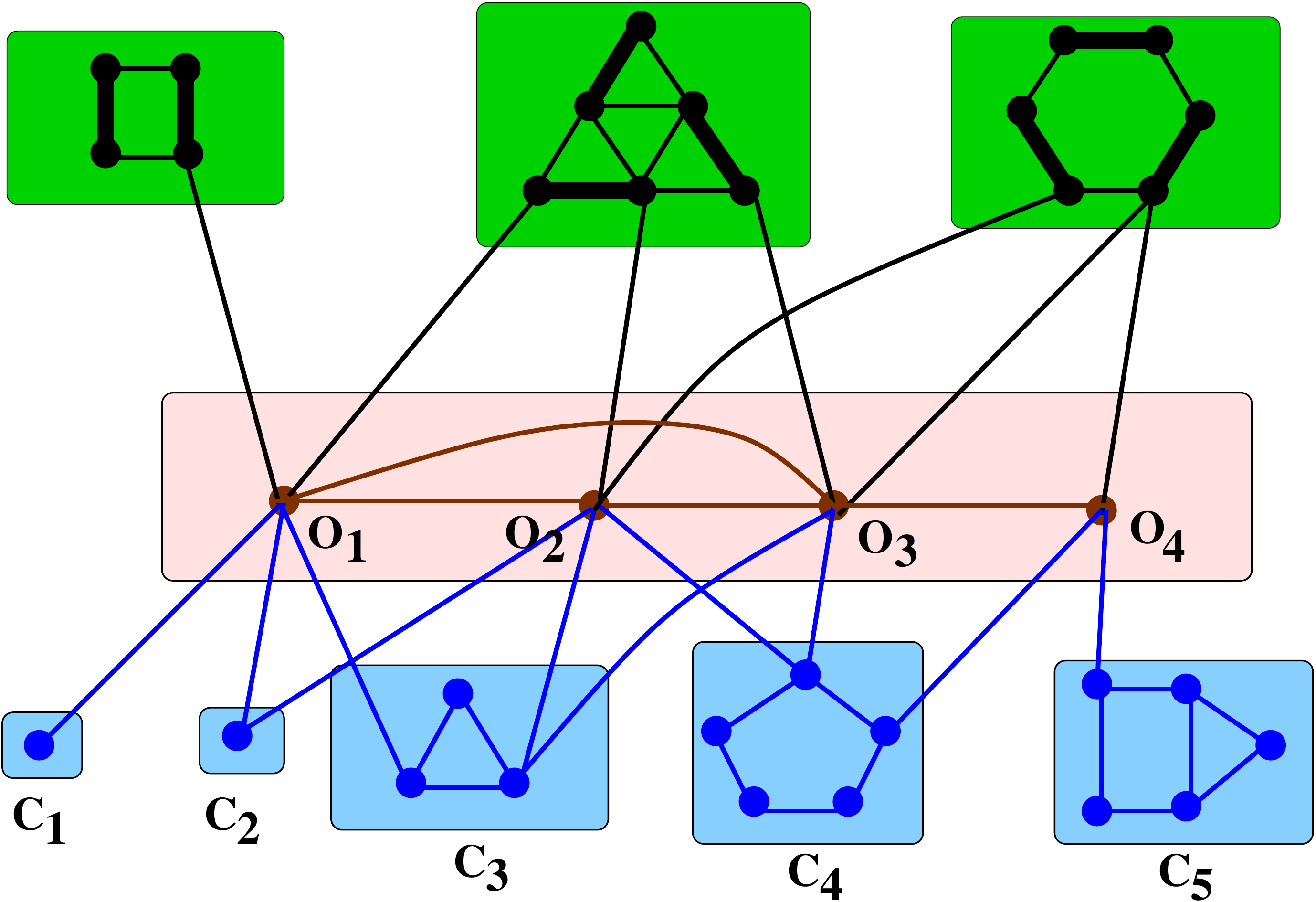}
\caption{A schematic of the Gallai-Edmonds decomposition of an arbitrary graph: Black vertices are the {\em unreachable} (u-type) vertices of the Gallai-Edmonds decomposition. They are always matched to an adjacent u-type vertex in any maximum matching. Brown vertices are the {\em odd} (o-type) vertices of the Gallai-Edmonds decomposition (labeled O). Links between two u-type vertices and between a u-type vertex and an o-type vertex are colored black. Links between two o-type vertices are colored brown. Blue vertices are the  {\em even} (e-type) vertices of the Gallai-Edmonds decomposition. Links between two e-type vertices and between an e-type vertex and an o-type vertex are colored blue. In any maximum matching, an e-type vertex is either matched with an adjacent vertex or remains unmatched, {\em i.e.} hosts a monomer. Each o-type vertex is always matched with an adjacent e-type vertex in any maximum matching. Note that an e-type vertex can never be adjacent to a u-type vertex. Upon deleting the links between the e-type and o-type vertices, the e-type vertices split up into connected {\em factor-critical} components (labeled C); each factor critical component has a perfect matching if any one of its vertices is deleted. Consequently, it hosts at most one monomer in any maximum matching of the whole graph, and this monomer can be on any of its vertices. Any o-type vertex always has links to more than one of these factor critical components. The total number of monomers in any maximum matching equals the difference between the number of factor critical components and the number of o-type vertices. See Sec.~\ref{sec:Lovaszlocal} for a detailed discussion.}
\label{fig:GallaiEdmonds}
 \end{figure}

\section{``Local'' version of the Lov{\'a}sz-Anderson result}
\label{sec:Lovaszlocal}
We now present a ``local'' argument that proves $Z=M$ and constructs a maximally-localized basis of wavefunctions for the topologically protected zero-energy eigenstates of a general Majorana network. The local aspect of our argument is this: We show that the entire graph associated with the real skew-symmetric matrix $a_{r r'}$ can be broken up into non-overlapping connected ``${\mathcal R}$-type'' regions ${\mathcal R}_\mu$ ($\mu=1,2...N_R$), such that the number of monomers ${\mathcal I}_\mu$ that must exist in any one such region ${\mathcal R}_\mu$ in any maximum matching equals the number of linearly-independent null vectors that are guaranteed to have all their nonzero amplitudes confined entirely within this region ${\mathcal R}_\mu$. Importantly, this remains true even when the graph as a whole forms a single connected component, {\em i.e.} each of these regions ${\mathcal R}_\mu$ remains connected to the rest of the graph. The fact that one can always construct a basis whose wavefunctions remain localized within individual ${\mathcal R}$-type regions in this generic case is thus a nontrivial property of the zero mode subspace.

\subsection{The Gallai-Edmonds decomposition}
\label{subsec:GallaiEdmonds}
\begin{figure}
 		\includegraphics[width=\columnwidth]{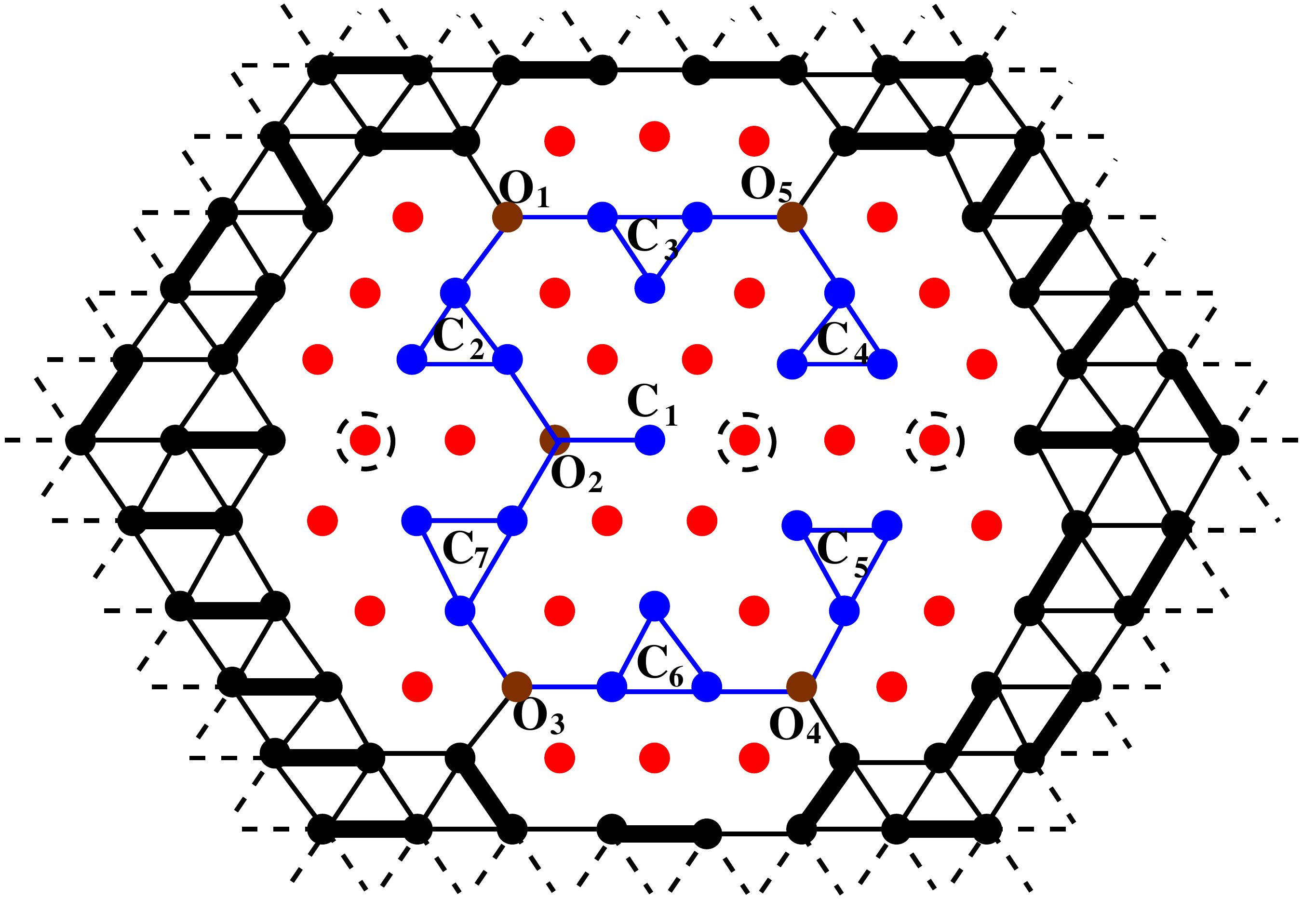}
\caption{A fragment of the triangular lattice, with dashed lines denoting links that connect it to the rest of the triangular lattice. Red denotes vacancies or deleted vertices, corresponding to missing Majorana modes. Other colors, labels and figure elements follow the conventions of Fig.~\ref{fig:GallaiEdmonds}.
Any maximum matching of the entire triangular graph (such a maximum matching is of course free to use the links connecting this fragment to the rest of the graph) is forced to leave two of the blue vertices unmatched and match each of the brown vertices with one of its blue neighbours, independent of the dimer configuration anywhere else in the lattice. Thus, any maximum matching has two
       monomers confined to the blue vertices of the ${\mathcal R}$-type region comprising the brown and blue vertices. Consequently, there are two linearly-independent topologically protected 
        collective Majorana excitations of $H_{\rm Majorana}$ (Eq.~\ref{eq:H_Majorana}) that are guaranteed to live entirely within this ${\mathcal R}$-type region, with support only on the blue vertices.  If any one of the three circled vacancies are removed, {\em i.e.} the corresponding localized Majorana mode reinstated, then the ${\mathcal R}$-type region will host exactly one monomer in any maximum matching, and there will be exactly one topologically protected Majorana excitation of $H_{\rm Majorana}$ supported on the blue vertices of the ${\mathcal R}$-type region. See Sec.~\ref{sec:Lovaszlocal} for a detailed discussion.}
\label{fig:TriangularRtype}
 \end{figure}
The Gallai-Edmonds decomposition~\cite{Gallai,Edmonds,Lovasz_Plummer} of an arbitrary graph starts with any one maximum matching of the graph and partitions all the vertices of the graph into three categories, even, odd, and unreachable. Even (e-type) vertices are those vertices that can be reached by at least one {\em alternating path} (consisting alternately of unmatched and matched edges) of even length ({\em i.e.}, with the total number of edges traversed being even) starting from some monomer of the maximum matching. Thus, unmatched vertices are themselves e-type, being reachable by a zero length alternating path. Unreachable (u-type) vertices are ones that cannot be reached by any alternating path from any monomer of the maximum matching. All other vertices that do not fall into either of these two categories are labeled odd. 

Although this partitioning is obtained by starting with any one maximum matching of the graph, it is actually unique, and represents a basic structural property of the graph itself~\cite{Lovasz_Plummer}. For the analysis presented here, the actual procedure for obtaining the e-type, o-type, or u-type label of each vertex is not important. It is enough to note that one can use efficient and well-studied combinatorial algorithms~\cite{Kececioglu_Pecqueur,Lovasz_Plummer} for finding one maximum matching and labeling the vertices in this manner. 
Indeed, as we see below, the importance of this labeling lies in the structural information it provides.

For instance, it is not hard to see that there can be no edges of the graph between e-type and u-type vertices~\cite{Lovasz_Plummer}. 
Further, the subgraph $G_{e}$ of e-type vertices and their mutual edges (obtained by deleting all u-type and o-type vertices and edges connecting them to the e-type vertices) splits into $N_{c}$ disjoint connected components $C_i$ ($i=1,2\dots N_c$). Each such $C_i$ has an odd number $B_i$ of e-type vertices~\cite{Lovasz_Plummer}, and is {\em factor critical}, {\em i.e.}, if any one vertex of $C_i$ is deleted, the resulting smaller subgraph with $B_i-1$ vertices has a perfect matching~\cite{Lovasz_Plummer}. 

Equivalently, maximum matchings of any such connected component $C_i$ have exactly one monomer, and this monomer can be on any vertex of $C_i$. Additionally, each o-type vertex is connected to more than one factor critical component by edges of the graph (if any o-type vertex was connected to only one factor critical component, all vertices of this factor critical component would always be matched in any maximum matching, contradicting the fact that these vertices are e-type). 

Turning to maximum matchings of the full graph, any maximum matching matches each u-type vertex with some other u-type vertex, while each o-type vertex is always matched to some e-type vertex in one of the factor critical components $C_i$~\cite{Lovasz_Plummer}. Two different o-type vertices are always matched to e-type vertices in distinct factor critical components. In any maximum matching, a given e-type vertex can either host a monomer, or be matched to another e-type vertex, or be matched to an o-type vertex~\cite{Lovasz_Plummer}. Each of the factor critical components $C_i$ hosts at most one monomer in any maximum matching of the full graph. 

Further, for any given factor critical component $C_i$, one can find a maximum matching that places a single monomer on this component, which, by the definition of factor criticality, can be on any vertex of this component. Finally, the total number of monomers in a maximum matching of the full graph is $2M=N_{c} - N_{o}$, where $N_{o}$ is the number of odd vertices of the full graph~\cite{Lovasz_Plummer}.

Finally, it is worth emphasizing that all these properties can be established by a sequence of elementary but somewhat involved combinatorial arguments~\cite{Gallai,Edmonds,Lovasz_Plummer} that do not need to invoke any results from linear algebra; for a particularly illuminating exposition along these lines, the reader is referred to the classic textbook of Lovasz and Plummer~\cite{Lovasz_Plummer}.

\subsection{Construction of maximally-localized basis of zero modes}
\label{subsec:Construction}

As before, let $2Z$ be the number of topologically protected null vectors of the $2L \times 2L$ real skew-symmetric matrix $a_{r r'}$, let $2M$ be the number of monomers in any maximum matching of the associated graph $G$, and let $N \equiv L-Z$.
The subgraph consisting of matched vertices (of any particular maximum matching) and all edges connecting any two of the matched vertices has a perfect matching, and therefore, by Tutte's theorem, the submatrix of $a_{r r'}$ associated with this subgraph has no topologically protected null vectors. Following the first part of the standard argument reproduced earlier, this implies the inequality Eq.~\ref{eq:inequality1}: $2Z \leq 2M$. We now give a prescription for constructing $2M$ linearly-independent topologically protected null vectors of $a_{r r'}$. This construction, in conjunction with the inequality $2Z \leq 2M$, constitutes our proof of Lov{\'a}sz and Anderson's result $Z=M$.

Consider first the matrix $a^{(C_i)}_{r r'}$,  which is defined as the restriction of the original skew-symmetric adjacency matrix $a_{r r'}$ to the vertices of $C_i$. We begin with the observation that $a^{(C_i)}_{r r'}$ has exactly one topologically protected zero mode. To see this, recall from the previous section that any maximum matching of $C_i$ has exactly one monomer, which can be at any site of $C_i$. Therefore, if we remove any one site of $C_i$, we are left with a subgraph with $B_i-1$ vertices which has a perfect matching. This implies that $a^{(C_i)}_{r r'}$ has a submatrix of size $B_{i} -1$ whose determinant does not vanish identically. Therefore the number $z_i$ of topologically protected null vectors of $a^{(C_i)}_{r r'}$ must satisfy $z_i \leq 1$ by our earlier argument. Since $a^{(C_i)}_{r r'}$ is a real skew-symmmetric matrix, its nonzero eigenvalues come in pure imaginary pairs $(i\lambda, -i \lambda)$. Since $B_i$, the dimension of $a^{(C_i)}_{r r'}$, is odd, this immediately implies that $z_i \geq 1$. Therefore $z_i=1$ for each of the $N_c$ factor critical components $C_i$. We denote the corresponding normalized real-valued null vector by $\rho^{(i)}(r)$, where $r$ ranges over all vertices of $C_i$. 

\iffalse
To begin, we delete all the u-type vertices and the edges connecting them to the o-type vertices, thus leaving behind a subgraph $G'$ consisting only of e-type and o-type vertices and edges between them. $G'$ has three kinds of edges: edges connecting two o-type vertices, edges connecting two e-type vertices, and edges connecting an o-type vertex to an e-type vertex. If we delete all edges connecting e-type vertices to o-type vertices, the Gallai-Edmonds decomposition guarantees that the subgraph $G_{e}$ of even vertices splits into $N_{c}$ disjoint connected components $C_i$ ($i=1,2\dots N_c$), each containing an odd number $B_i$ of vertices.

\fi

%Next we delete all the edges between two o-type vertices to obtain the graph $G''$. This reduced subgraph $G''$ has two types of edges: edges between two e-type vertices, which we color blue, edges connecting an e-type vertex to an o-type vertex, which we color red. If we  remove all red edges of $G_{eo}$, we know from the properties of the Gallai-Edmonds decomposition that the subgraph $G_{e}$ of even vertices and blue edges splits into $N_{c}$ disjoint connected components $C_i$ ($i=1,2\dots N_c$), each containing an odd number $B_i$ of vertices.

We now construct $2M$ linearly-independent superpositions of these $N_{c}$ wavefunctions $\rho^{(i)}(r)$ (note that $\rho^{(i)}$ and $\rho^{(j)}$ have non-overlapping supports whenever $i \neq j$).
To this end, we start by defining an auxillary bipartite graph $G''$ as follows: Start with the original graph and replace each factor critical component $C_i$ by a single vertex $c_i$, delete all edges connecting o-type vertices to each other, and delete all u-type vertices as well as edges incident on them. If a given o-type vertex $o_k$ has at least one edge connecting it to some vertex $r \in C_i$ in the original graph, $o_k$ has a single edge connecting it to the corresponding vertex $c_i$ of $G''$. Thus, all the edges which connect any of the vertices of a given $C_i$ to a given o-type vertex in the original graph are ``collapsed'' in the bipartite graph $G''$ into a single edge connecting the corresponding vertex $c_i$ to this o-type vertex. 

We define the skew-symmetric bipartite adjacency matrix $\tilde{a}$, whose nonzero elements correspond to these edges of $G''$, as follows:
\begin{eqnarray}
\tilde{a}_{o_k c_i} &=& \sum_{r \in C_i} a_{o_k r} \rho^{(i)}_r \; ,\nonumber \\
\tilde{a}_{c_i o_k} & = & - \tilde{a}_{o_k c_i} \; , 
\label{eq:Defnnewadjacencymatrix} 
\end{eqnarray}
where the sum over $r \in C_i$ receives contributions from all vertices of $C_i$ that had a link to the o-type vertex $o_k$ in the original graph. 

Clearly, the bipartition of $G''$ assigns all $N_c$ vertices $c_i$ to one sublattice (which we declare to be the
 $A$ sublattice) and the $N_o$ o-type  vertices of the original graph to the opposite sublattice, which we label
  the $B$ sublattice.  Any matching of $G''$ must therefore have a minimum of $N_c-N_o$ monomers, since there are
   $N_c$ $A$-sublattice sites and $N_o < N_c$ $B$-sublattice sites. To obtain a maximum matching of $G''$ that has exactly $N_c - N_o$ monomers, we can start with a maximum matching
    of the original graph and make the following construction: Match each $B$-sublattice vertex $o_k$ of $G''$ to the $A$-sublattice site $c_j$ chosen to correspond to the factor critical component $C_j$ into which $o_k$ is matched by the maximum matching of the original graph. This gives us a maximum matching of $G''$ which matches all $B$-sublattice sites and has exactly $N_c-N_o$ monomers living on $A$-sublattice sites. For any particular factor critical component $C_i$, the Gallai-Edmonds decomposition guarantees that there exists a maximum matching of the full graph which places a single monomer on this $C_i$. Our construction maps this to a maximum matching of $G''$ which places a monomer on the corresponding $A$-sublattice site $c_i$.

In the Dulmage-Mendelsohn decomposition of $G''$ (which is just the bipartite version of the Gallai-Edmonds decomposition, in which the role of the factor critical components is played by single e-type sites), all $A$-sublattice sites of $G''$ are thus seen to be e-type, all $B$-sublattice sites of $G''$ are o-type, and there are no u-type sites. Since $G''$ has no u-type sites, and no edges between any two o-type sites, the arguments of Ref.~\cite{Bhola_etal_PRX} imply that it must be made up of $N_R \geq 1$ disjoint connected components, {\em each of which has more $A$-sublattice sites than $B$-sublattice sites}. This is of course trivially true if $N_R=1$, since $G''$ as a whole does have more $A$ sites than $B$ sites; the key point is that this is guaranteed to be true ``locally'', at the level of individual connected components of $G''$. 

Following the terminology of Ref.~\cite{Bhola_etal_PRX}, 
we denote these components ${\mathcal R}_{A}^{(\mu)}$ ($\mu=1,2 \dots N_R$). 
The excess of $A$ sites over $B$ sites in ${\mathcal R}_{A}^{(\mu)}$ is referred to as the imbalance ${\mathcal I}_\mu$; thus, there are $m^{(\mu)}_A$ $A$-sublattice sites and $m^{(\mu)}_B$ $B$-sublattice sites in ${\mathcal R}_A^{(\mu)}$, with ${\mathcal I}_\mu = m^{(\mu)}_A - m^{(\mu)}_B$. In the language of Ref.~\cite{Bhola_etal_PRX}, these are the ${\mathcal R}_A$-type regions of $G''$. Indeed, unlike the general bipartite case discussed in Ref.~\cite{Bhola_etal_PRX}, $G''$, by virtue of the particulars of its construction, only has ${\mathcal R}_A$-type regions; in the language of Ref.~\onlinecite{Bhola_etal_PRX}, it has no ${\mathcal R}_B$ type regions or ${\mathcal P}$-type regions.

Each ${\mathcal R}_{A}^{(\mu)}$ hosts exactly ${\mathcal I}_\mu$ monomers on its $A$-sublattice sites in any maximum matching of $G''$. The total number of monomers in any maximum matching of $G''$ is thus $\sum_{\mu=1}^{N_R} {\mathcal I}_\mu$.
From the definiton of ${\mathcal I}_{\mu}$, we see that this sum equals $N_c -N_o$, which, by the Gallai-Edmonds decomposition, gives the number of monomers $2M$ in any maximum matching of the original graph, as it of course must. Thus we have
\begin{eqnarray}
2M &=& \sum_{\mu=1}^{N_R} {\mathcal I}_\mu \; .
\end{eqnarray}

Each ${\mathcal R}_A^{(\mu)}$ corresponds in a natural way to a subgraph ${\mathcal R}_{\mu}$ of the original graph. To obtain this from ${\mathcal R}_A^{(\mu)}$, we expand out each $c_i \in {\mathcal R}_A^{(\mu)}$ to recover all the e-type vertices of the corresponding factor critical component $C_i$ and the edges incident on these e-type vertices. In addition, we reinstate all the edges of the original graph between any two o-type sites that both belong to ${\mathcal R}_A^{\mu}$. Clearly, each such ${\mathcal R}_\mu$ hosts exactly ${\mathcal I}_\mu$ monomers in any maximum matching of the original graph.

Next, we note that a ${\mathcal R}_A$-type region ${\mathcal R}_A^{(\mu)}$ hosts exactly ${\mathcal I}_\mu$ topologically protected zero modes of $\tilde{a}_{o_k c_i}$, which have nonzero amplitude only on the $A$-sublattice sites of ${\mathcal R}_A^{(\mu)}$. To see this, note that the any such zero mode has to satisfy $m^{(B)}_\mu$ equations in $m^{(A)}_\mu$ variables, leading to ${\mathcal I}_\mu = m^{(A)}_\mu - m^{(B)}_\mu$ topologically protected zero mode solutions. Thus, $\tilde{a}$ has $\sum_{\mu} {\mathcal I}_\mu = 2M$ topologically protected zero modes, with ${\mathcal I}_\mu$ such linearly-independent modes coexisting in region ${\mathcal R}_A^{(\mu)}$. Let us denote these null vectors of $\tilde{a}$ by
$v^{(\mu)}_\alpha $ with $\alpha=1,2\dots {\mathcal I}_\mu$, so that the component of $v^{(\mu)}_\alpha $ on some $A$-sublattice site $c_i \in {\mathcal R}_A^{(\mu)}$ is denoted as $v^{(\mu)}_\alpha (c_i)$.

It only remains to use this result to obtain the corresponding zero mode wavefunctions of the original skew-symmetric adjacency matrix $a_{rr'}$. This is done as follows: For each of the factor critical components $C_i$ that correspond to $c_i \in {\mathcal R}_A^{(\mu)}$, we recall that we have at our disposal the normalized zero mode wavefunctions $\rho^{(C_i)}(r)$ of $a^{(C_i)}_{r r'}$, the restriction of $a_{r r'}$ to the factor critical component $C_i$. Each of these $\rho^{(C_i)}(r)$ are nonzero only on sites $r \in C_i$. Using these, we form ${\mathcal I}_\mu$ linearly-independent wavefunctions defined on e-type vertices $r \in {\mathcal R}_\mu$:
\begin{eqnarray}
\phi^{(\mu)}_{\alpha}(r) = \sum_{c_i \in {\mathcal R}_{A}^{(\mu)}} v^{(\mu)}_{\alpha}(c_i) \rho^{(i)}(r) \; \; ({\rm with} \; \alpha = 1,2 \dots {\mathcal I}_\mu) \nonumber \\
\label{eq:construction}
\end{eqnarray}
\begin{figure}
 		\includegraphics[width=\columnwidth]{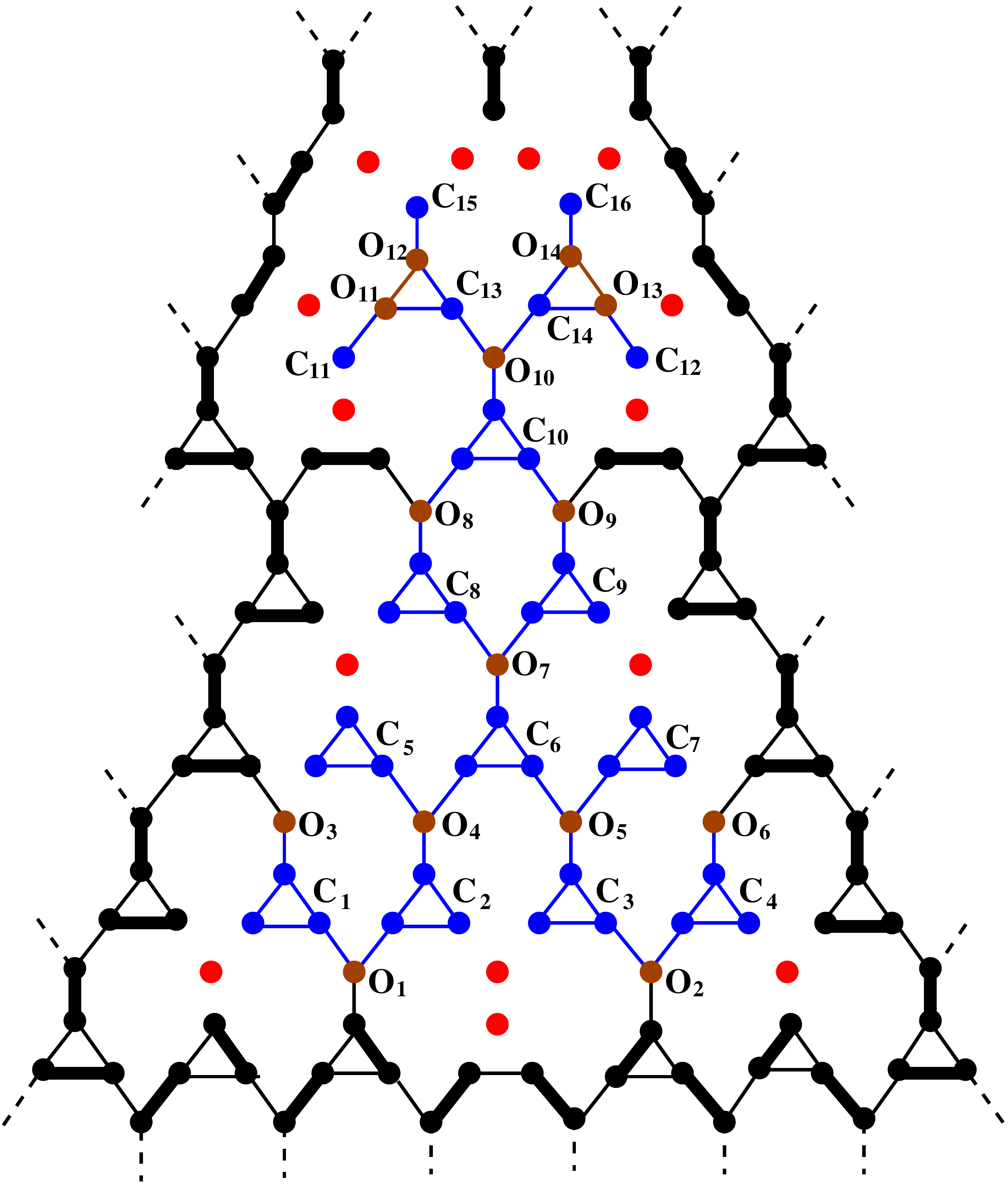}
\caption{A fragment of the star-triangle lattice studied in Ref.~\onlinecite{JFu_PRB} in the context of a Kitaev-type model that exhibits a chiral spin liquid phase. Dashed links denote connections to the rest of the network graph. Red denotes vacancies or deleted vertices, corresponding to missing Majorana modes. Other colors, labels, and figure elements follow the conventions of Fig.~\ref{fig:GallaiEdmonds}. Any maximum matching of the entire graph (such a maximum matching is of course free to use the links connecting this fragment to the rest of the graph) is forced to leave two of the blue vertices unmatched. Thus, any maximum matching has two monomers confined to the blue vertices of the ${\mathcal R}$-type region comprising blue and brown vertices.  Consequently, there are two linearly-independent topologically protected 
        collective Majorana excitations of $H_{\rm Majorana}$ (Eq.~\ref{eq:H_Majorana}) that are guaranteed to live entirely within this ${\mathcal R}$-type region, with support only on the blue vertices.}
\label{fig:startriangleRtype}
 \end{figure}
Since each $ \phi^{(\mu)}_{\alpha}(r)$ is zero on all o-type sites of the original graph, and since u-type sites only have edges connecting them to o-type sites in the original graph,  the eigenvalue equation for zero modes of $a_{r r'}$ is trivially satisfied by each $ \phi^{(\mu)}_{\alpha}(r)$ (for $\alpha=1,2,\dots {\mathcal I}_\mu$ for each $\mu=1,2\dots N_R$) on all u-type sites of the original graph. On all o-type sites, the zero-energy eigenvalue equation for $a_{r r'}$ is satisfied by virtue of fact that the $v^{(\mu)}_{\alpha}$ are null vectors of $\tilde{a}_{o_k c_i}$, whose matrix
 elements depend on $\rho^{(i)}$ in a way that ensures that these eigenvalue equations are automatically satisfied. And crucially, the zero
  mode eigenvalue equation for $a_{r r'}$ is also seen to be satisfied on all e-type sites sites of the graph by virtue of the fact that $\rho^{(i)}$ is a topologically protected zero mode of $a^{(C_i)}_{rr'}$. 

We have thus constructed ${\mathcal I}_\mu$ linearly-independent zero modes of $a_{r r'}$ that have nonzero amplitudes only on the e-type sites in the region ${\mathcal R}_\mu$ of the original graph (for each $\mu=1,2,\dots N_R$). Since $\sum_{\mu} {\mathcal I}_\mu = 2M$ (where $2M$ is the number of monomers in any maximum matching of the graph associated with $a_{r r'}$), and we have already argued that $2Z \leq 2M$ (where $2Z$ is the number of topologically protected zero modes of $a_{r r'}$), this linearly-independent set forms a basis for the topologically protected null space of $a_{r r'}$.

Notice that our basis is maximally-localized in the following sense: Each basis vector is supported entirely on e-type sites within one particular region ${\mathcal R}_\mu$. This is independent of the actual values of the nonzero matrix elements of $a_{r r'}$. Additionally, these nonoverlapping regions ${\mathcal R}_\mu$ are also independent of these actual values of the nonzero matrix elements. An easy consequence is that the basis-independent on-shell zero-energy Green function $G(r,r')$, defined as
\begin{eqnarray}
G(r,r') &\equiv& \sum_{\mu=1}^{N_R} \sum_{\alpha = 1}^{{\mathcal I}_\mu} \phi^{(\mu)}_{\alpha}(r) \phi^{(\mu)}_{\alpha}(r') \; ,
\end{eqnarray}
is guaranteed to be zero unless $r$ and $r'$ both belong to any one region ${\mathcal R}_\mu$, and this property is independent of the actual numerical values of the nonzero matrix elements of $a_{ r r'}$. This is clearly the strongest topologically protected statement one can make about the localization properties of the zero energy Green function, and it follows directly from the localization properties of our basis. In this sense, our construction yields a maximally localized basis of topologically protected Majorana fermion excitations of the network as a whole.

\section{Algorithmic considerations}
\label{sec:Algorithmic}
We now discuss some algorithmic issues of stability and accuracy that are likely to arise when using our local version of the Lov{\'a}sz-Anderson result to obtain a maximally-localized basis of zero modes for large random networks. These issues have their origin in the bipartite nature of the graph $G''$ used in our construction. As is well-understood from previous studies of random-hopping models with bond disorder and site dilution~\cite{Sanyal_Damle_Motrunich_PRL,Sanyal_etal_SU2,Motrunich_Damle_Huse_GadeWegnerPRB,Willans_Chalker_Moessner_PRB} the corresponding bipartite adjacency matrix $\tilde{a}$ is expected to have a large pile-up of eigenstates near the center of the band in the two-dimensional case. When the size of the problem is large, as is the case when individual ${\mathcal R}^{(\mu)}_A$ regions in our construction are large and host a large number of zero modes, this is expected to lead to stability issues when it comes to constructing the topologically protected zero modes that are tied to the band center itself. 

This effect also makes the study of zero mode wavefunctions challenging on slightly-diluted bipartite lattices. Our proposed solution to this potential problem relies on a simple but useful observation~\cite{Bhola_Biswas_Damle_wavefunction}: Consider any one region ${\mathcal R}^{(\mu)}_A$
hosting ${\mathcal I}_\mu$ zero modes in any maximum matching of the bipartite graph $G''$. Choose any particular maximum matching, with monomers located on sites $c_k$ ($k=1,2 \dots {\mathcal I}_\mu$). If we delete all the $c_k$ except one, say $c_{j}$, it is clear that all other ${\mathcal R}^{(\nu)}_A$ with $\nu \neq \mu$ remain unchanged in the Dulmage-Mendelsohn decomposition of the truncated graph $G''_{j}$. On the other hand, ${\mathcal R}^{(\mu)}_A$ splits in general into a truncated ${\mathcal R}$-type region ${\mathcal R}_A^{j}$ that hosts exactly one monomer in any maximum matching of $G''_{j}$, and one or more ${\mathcal P}$-type regions that are perfectly matched in any such maximum matching.

This observation suggests the following ``divide-and-conquer'' algorithm which has also been explored in the context of the bipartite quantum percolation problem~\cite{Bhola_Biswas_Damle_wavefunction}: Start with a particular maximum matching of ${\mathcal R}^{(\mu)}_A$ as described above, 
with ${\mathcal I}_\mu$ monomers on sites $c_k$ ($k=1,2 \dots {\mathcal I}_\mu$) in region ${\mathcal R}^{(\mu)}_A$. 
For each $j=1,2 \dots {\mathcal I}_\mu$, we implement the following procedure:
Delete all sites $c_k \neq c_j$  and obtain the truncated ${\mathcal R}$-type region ${\mathcal R}_A^{j}$ with 
$N_{\mu}^{(j)}$ $A$-sublattice sites and $N_{\mu}^{(j)}-1$ $B$-sublattice sites.
To obtain the unique (up to an overall scale) zero mode supported on the truncated 
${\mathcal R}$-type region ${\mathcal R}_A^{j}$. we set the wavefunction amplitude at $c_j$ to unity. With this in hand, compute the $N_{\mu}^{(j)}-1$ other components of the wavefunction by solving the system of $N_{\mu}^{(j)}-1$ Schrodinger equations for these $N_{\mu}^{(j)}-1$ variables using some stabilized version of Gaussian elimination. 

When the iteration over $j=1,2 \dots {\mathcal I}_\mu$ is complete, this procedure gives ${\mathcal I}_\mu$ zero modes confined
to the region ${\mathcal R}^{(\mu)}_A$, as required. 
Note that this algorithm guarantees that these ${\mathcal I}_\mu$ wavefunctions are linearly-independent to arbitrary accuracy because only one of them has nonzero amplitude at any particular $c_k$.

\section{Discussion}
\label{sec:Discussion}
Given that the foregoing approach leads to such robust conclusions regarding vacancy effects in Majorana networks, it is of some interest to ask what conclusions (if any) can be drawn from such arguments about free-fermion systems modeled by a tight-binding model of canonical fermions on a general lattice.   In addition, with our construction now in hand, it is interesting to revisit the original proof of Ref.~\cite{Anderson} and contrast our approach with that of Ref.~\cite{Anderson}. Further, it is also useful to illustrate the ideas developed here with some particularly simple examples in which the veracity of our conclusions may be directly verified ``by hand''. The discussion below addresses each of these in turn,  and then conncludes by sketching a heuristic argument that relates the  ${\mathcal R}$-type regions of a slightly-diluted non-bipartite lattice to various components of the Dulmage-Mendelsohn decomposition of a ``parent'' bipartite lattice.

\subsection{Vacancy-effects in tight-binding models of canonical fermions}

Consider a tight-binding model for free canonical fermions, with a canonical fermion orbital of energy $V_r$ at each surviving vertex $r$ of a general graph (in which some vertices have been removed to model the effects of vacancies), and complex hopping amplitudes $t_{\langle r r'\rangle}$ defined on links $\langle r r' \rangle$ that connect surviving vertices of the graph. The Hamiltonian for the canonical fermions can be written as
\begin{eqnarray}
H_{\rm canonical} &=& \sum_{r, r' } T_{ r r'}f^{\dagger}_{r} f_{r'}  \, ,
\end{eqnarray}
where $T_{r r'} = t_{\langle r r'\rangle} + \delta_{rr'} V_r/2$ and $T_{r'r} = T^{*}_{r r'}$.

\iffalse
\begin{eqnarray}
H_{\rm canonical} &=& \sum_{\langle r r' \rangle }( t_{\langle r r'\rangle }f^{\dagger}_{r} f_{r'} + h.c.) + \sum_r V_r f^{\dagger}_{r} f_{r} \, .
\end{eqnarray}
\fi

Defining $f_r = (a_r + i b_r)/2$, where $a_r$ and $b_r$ are Majorana fermion operators, this can be rewritten as a Majorana network Hamiltonian for a bilayer version of the original graph
\begin{eqnarray}
H_{\rm canonical} &=& \frac{i}{4} \sum_{\alpha,\beta} \eta_{r \alpha} ({\mathcal I}_{r r'} \otimes {\mathbf 1}_{\alpha \beta} + {\mathcal R}_{r r'} \otimes i\sigma^{y}_{\alpha \beta} )\eta_{r' \beta} \; , \nonumber \\
&&
\end{eqnarray}
where $\alpha$ and $\beta$ take on values $a, b$ in the summation over these layer indices, and we define $\eta_{r a} = a_r$, $\eta_{r b} = b_r$. Here, ${\mathbf 1}$ and $i \sigma^{y}$ respectively are the $2 \times 2$  identity and Pauli matrices acting in the layer space, ${\mathcal I}_{r r'}$ is an {\em antisymmetric} matrix defined by ${\mathcal I}_{r r'} = {\rm Im}(T_{r r'})$ and ${\mathcal R}_{r r'}$ is a {\em symmetric} matrix defined by ${\mathcal R}_{r r'} = {\rm Re}(T_{r r'}) $.

From this expression, it is evident that the tight-binding model for canonical fermions on a general graph with one orbital on each vertex of the graph reduces to two {\em independent} copies of a Majorana network Hamiltonian on the same graph if and only if there are no onsite energy terms and the hopping amplitudes are purely imaginary. In this case, our arguments go through and give us a detailed characterization of the topologically protected zero energy states that are tied to the Fermi energy of $H_{\rm canonical}$. Following the approach outlined above, this can be generalized slightly to include pure-imaginary pairing amplitudes as well. 

However, we caution that the general case requires us to directly analyze the bilayer graph. In the general case with arbitrary complex $t_{\langle r r'\rangle}$ and a nonzero site energy $V_r$, a missing site in the original tight-binding model maps to a pair of deleted vertices in a bilayer graph in which both layers are nontrivially coupled to each other. In this case, it is not clear if pair dilution can lead to any topologically protected localized modes of the type we have discussed here. 

The only exceptions are topologically protected zero modes in particle-hole symmetric tight-binding models with arbitrary complex hopping amplitudes, {\em i.e.} tight-binding models on bipartite lattices with $V_r = 0$. The origin of these modes has been discussed extensively in earlier work~\cite{Bhola_etal_PRX}, and we learn nothing new by reformulating that discussion using the approach developed here.

\subsection{Revisiting Anderson's approach}

Ref.~\cite{Anderson} starts with any one arbitrarily chosen maximum matching of the associated graph, with monomers at vertices $r_l$, with $l \in (1,2\dots 2M)$. For any one vertex $r_k$ chosen arbitrarily from these unmatched vertices $\{r_l \}$, Ref.~\cite{Anderson} proves that there must exist a null vector of $ia_{r r'}$, which has nonzero amplitude at $r_k$ and amplitude equal to zero at all other unmatched vertices.  Since this is true independently for each such $r_k$, this establishes the presence of $2M$ linearly independent null vectors. 

In contrast, our local approach uses the fact that ${\mathcal R}$-type regions can be identified directly from the Gallai-Edmonds decomposition of the associated graph, without the need for any numerical calculation, and goes on to establish the existence of a certain number of null vectors that live entirely within each ${\mathcal R}$-type region. Thus, it incorporates at the very outset the topologically protected localization properties of the zero-energy on-shell Green function $G(r,r')$ (which follow from the fact that $G(r,r')$, when evaluated using the basis constructed by our approach, is obviously zero unless $r$ and $r'$ lie in the same ${\mathcal R}$-type region). If the ${\mathcal R}$-type regions are small in extent, this immediately provides a strict and useful upper bound on the localization length $\xi_G$ of $G$, which must be bounded above by the linear dimension of the largest ${\mathcal R}$-type region. Unlike in our construction, these properties do not emerge explicitly in any straightforward way from the arguments of Ref.~\cite{Anderson}; in that approach, they would therefore need to be discovered separately by explict numerical computation.

%\subsection{Some examples}
%\label{subsec:Examples}

\subsection{Simple illustrative examples}
The local nature of our approach, which distinguishes it from earlier work, is best illustrated by some simple examples involving small ${\mathcal R}$-type regions, in which we can work everything out ``by hand''. It is interesting to do this first on the site-diluted triangular lattice, since a triangular lattice of localized Majorana modes~\cite{Laumann_Ludwig_Huse_Trebst} serves as a model for the Majorana network associated with a vortex lattice in time-reversal symmetry breaking $p$-wave superconductors~\cite{RBiswas_PRL,Read_Green,Ivanov}.

Fig.~\ref{fig:TriangularRtype} shows a simple ${\mathcal R}$-type region consisting of twenty-four sites,  formed due to a clustering of vacancies. This region of the triangular lattice has seven factor critical components, six of them with three e-type vertices each, and one of them being made up of just a single e-type vertex. These factor criticalcomponents are connected to five o-type vertices, four of which connects this ${\mathcal R}$-type region to the rest of the lattice. Independent of the configuration of dimers in the rest of the lattice, this ${\mathcal R}$-type region necessarily hosts two monomers in any maximum matching of the full lattice. Our contruction shows that it also hosts two topologically protected zero modes that have nonzero amplitudes on the e-type sites of this region. Although the existence of these zero modes follows from purely local considerations, nonzero energy excitations cannot be determined by purely local considerations since the local Majorana modes within this ${\mathcal R}$-type region remains connected to the rest of the Majorana network by nonzero mixing amplitudes.

\iffalse
From this example and others like it, it appears that a clustering of vacancies is essential for seeding such ${\mathcal R}$-type regions on the diluated triangular lattice. In fact, this intuition has been put on a firm footing by Anstee and Tseng: under mild restrictions on the shape of the boundary of the sample, their analysis shows that a diluted triangular lattice always has a perfect matching if no two vacancies are within three lattice units of each other. 
%This is in sharp contrast to the diluted square lattice. In the case of a bipartite square lattice with $V$ sites and side-length $\sqrt{V}$, it is known that a perfect matching is only guaranteed if no two vacancies lie within ${\mathcal O}(V^{1/4})$ lattice separations of each other.
\fi

A Hamiltonian of the form Eq.~\ref{eq:H_Majorana} also describes the low-energy physics of Kitaev-type models of Majorana spin liquids and mean field theories for Majorana spin liquids~\cite{Kitaev_anyon,Yao_Lee,Yao_Kivelson_PRL,Chua_Yao_Fiete_PRB,Lai_Motrunich_PRB,MotomehypononagonKitaev,Sanyal_etal_SU2,JFu_PRB,Biswas_Fu_Laumann_Sachdev}
Motivated by this, Fig.~\ref{fig:startriangleRtype} displays another example of ${\mathcal R}$-type region, this time of relevance to the effect of vacancies in the Kitaev-like model on the star-triangle lattice or wine glass lattice~\cite{JFu_PRB}. In this case, the ${\mathcal R}$-type region shown is formed by the clustering of fourteen vacancies. It has sixteen factor critical components, ten of them consisting of three vertices that form a triangle, and four of them being single e-type sites. These factor critical components are connected to fourteen o-type sites, four of which connect this region to the rest of the lattice. Again, independent of the dimer configuration on the rest of the lattice, this ${\mathcal R}$-type region must host two monomers in any maximum matching of the full lattice, and there are correspondingly two zero modes with amplitudes on the e-type sites of the region.

\subsection{Outlook}

Finally, it is worth emphasizing that these examples, which were chosen for their simplicity, symmetry, and ease of visualization, are very unlikely to be typical of ${\mathcal R}$-type regions that actually arise in large random Majorana networks. Understanding the random geometry of such typical regions is clearly an interesting problem its own right. Although this computationally intensive problem is outside the scope of the present study, we close our discussion by outlining a simple heuristic picture that may serve as additional motivation for future work along these lines.

 The basis for our heuristic picture is the linear stability analysis of Ref.~\onlinecite{Bhola_etal_PRX}, which identifies topologically protected collective Majorana zero modes of bipartite Majorana networks that are perturbatively stable to additional next-nearest-neighbor couplings whose inclusion destroys the bipartiteness of the original network.
 The key ingredient in this analysis is the Dulmage-Mendelsohn decomposition of the parent bipartite network into ${\mathcal R}_A$-type, ${\mathcal R}_B$ type and ${\mathcal P}$-type regions~\cite{Bhola_etal_PRX}, with each ${\mathcal R}$-type region hosting a nonzero number ${\mathcal I}$ of topologically protected collective Majorana modes of the bipartite network. Here, ${\mathcal I}$ is the sublattice imbalance within the ${\mathcal R}$-type region, {\em i.e.} the modulus of the difference between the number of $A$-sublattice sites and the number of $B$-sublattice sites in the region.  
 
 Within leading-order perturbation theory in the additional non-bipartite couplings~\cite{Bhola_etal_PRX} that couple next-nearest-neighbors on the ``parent'' bipartite lattice, the argument given in Ref.~\onlinecite{Bhola_etal_PRX} shows that ${\mathcal R}$-type regions with odd ${\mathcal I}$ host a single topologically protected collective Majorana mode of the modified network, while ${\mathcal R}$-type regions with even ${\mathcal I}$ have no Majorana modes that survive the inclusion of these additional non-bipartite couplings.
 
 With this background in hand, we now  sketch our heuristic picture for the ${\mathcal R}$-type regions of site-diluted nonbipartite lattices. For concreteness, we focus our discussion below on the case of a site-diluted triangular lattice, which we view in this context as a square lattice with additional next-nearest-neighbor couplings along one diagonal of the parent square lattice. The perturbative stability argument given in Ref.~\cite{Bhola_etal_PRX} then suggests that the collective Majorana modes of the diluted triangular lattice can be ascribed to ${\mathcal R}$-type regions with odd ${\mathcal I}$ in the parent square lattice. At low dilution, ${\mathcal R}$-type regions of diluted square lattice become very big~\cite{Bhola_etal_PRX}, with their linear size scaling roughly as $n_{\rm vac}^{-5}$~\cite{Bhola_etal_PRX}, where $n_{\rm vac}$ is the site-dilution probability (density of vacancies); this is associated with an incipient Dulmage-Mendelsohn percolation phenomenon in the $n_{\rm vac} \to 0$ limit~\cite{Bhola_etal_PRX}. 
 
 In a large sample  of the diluted square lattice in this low-$n_{\rm vac}$ regime, there are typically two dominant ${\mathcal R}$-type regions; one of them is an ${\mathcal R}_A$-type region, while the other is an ${\mathcal R}_B$-type region~\cite{Bhola_etal_PRX}. Each of these two largest ${\mathcal R}$-type regions have an odd imbalance ${\mathcal I}$ with probability close to $0.5$~\cite{Bhola_etal_PRX}, and will therefore survive as an ${\mathcal R}$-type region of the triangular lattice with probability close to $0.5$ within this leading-order treatment of the additional couplings. This already suggests that the actual ${\mathcal R}$-type regions obtained from the Gallai-Edmonds decomposition of the triangular lattice (using the procedure given in this work) will have a nonzero probability for being very large in size in the low-dilution limit.
 
 Going beyond this leading-order picture, we see that the additional non-bipartite couplings can also have other effects that are not captured at leading order:  An ${\mathcal R}_A$-type region with odd ${\mathcal I}$ on the parent square lattice and another neighboring ${\mathcal R}_B$-type region with odd ${\mathcal I}$ on the parent square lattice can annihilate pair-wise, forming a larger ${\mathcal P}$-type region of the triangular lattice. Another possibility is that two neighboring ${\mathcal R}$-type regions of the square latice, one with odd imbalance ${\mathcal I}$ and the other with an even ${\mathcal I}$, merge to form a bigger ${\mathcal R}$-type region of the triangular lattice. 
 
The actual statistical properties of ${\mathcal R}$-type regions of a slightly-diluted triangular lattice are thus expected to be determined by a combination of these processes of annihilation, merging, and pair-wise annihilation of ${\mathcal R}$-type regions of the parent square lattice. Clearly this heuristic picture falls short of making precise predictions for these statistical properties. But it does suggest that the random geometry of these ${\mathcal R}$-type regions on the triangular lattice is likely to be very interesting in the low-dilution limit.

  \section{Acknowledgements}
  \label{sec:Acknowledgements}
  
  I gratefully acknowledge stimulating discussions with T.~Kavitha, J. Radhakrishnan, and D.~Sen, and fruitful collaborations on closely related earlier work (Refs.~\onlinecite{Sanyal_Damle_Motrunich_PRL,Sanyal_etal_SU2,Bhola_etal_PRX}) with R.~Bhola, S.~Biswas, J.~T.~Chalker,
  R.~Moessner, O.~I.~Motrunich, and S.~Sanyal. I am also grateful for the generous research support at the Tata Institute of Fundamental Research (TIFR) provided by DAE, India and in part by a J.C. Bose Fellowship  (JCB/2020/000047) of SERB, DST India, and by the Infosys-Chandrasekharan Random Geometry Center (TIFR).

\end{document}